# Suppression of Gate Screening on Edge Magnetoplasmons by Highly Resistive ZnO Gate

N. Kumada,[1] N.-H. Tu,[1] K.-i. Sasaki,[1,2] T. Ota,[1] M. Hashisaka,[1] S. Sasaki,[1] K. Onomitsu,[1] and K. Muraki[1]

[1]*NTT Basic Research Laboratories, NTT Corporation, 3-1 Morinosato-Wakamiya, Atsugi, 243-0198, Japan*

[2]*NTT Research Center for Theoretical Quantum Physics, NTT Corporation, 3-1 Morinosato-Wakamiya, Atsugi, 243-0198, Japan*

We investigate a way to suppress high-frequency coupling between a gate and low-dimensional electron systems in the gigahertz range by measuring the velocity of edge magnetoplasmons (EMPs) in InAs quantum Hall systems. We compare the EMP velocity in three samples with different electromagnetic environments—one has a highly resistive zinc oxide (ZnO) top gate, another has a normal metal (Ti/Au) top gate, and the other does not have a gate. The measured EMP velocity in the ZnO gate sample is one order of magnitude larger than that in the Ti/Au gate sample and almost the same as that in the ungated sample. As is well known, the smaller velocity in the Ti/Au gate sample is due to the screening of the electric field in EMPs. The suppression of the gate screening effect in the ZnO gate sample allows us to measure the velocity of unscreened EMPs while changing the electron density. It also offers a way to avoid unwanted high-frequency coupling between quantum Hall edge channels and gate electrodes.

High-frequency charge manipulation in the gigahertz (GHz) range in low-dimensional electron systems has been extensively investigated for a variety of objectives, such as quantum information processing [1,2], single-electron sources [3-6], electronic quantum optics [7-9], and nonreciprocal microwave devices [10-12]. Meanwhile, high-frequency charge dynamics in quantum Hall (QH) edge channels (ECs) and helical ECs in a two-dimensional topological insulator has been investigated to obtain information on the EC structure [13-18] and detect Tomonaga-Luttinger liquid nature of one-dimensional systems [19-26]. For these experiments, gate electrodes, which control electron density and define mesoscopic circuits, are indispensable. However, gate electrodes have side effects. Dephasing and dissipation of high-frequency charges are induced through the capacitive coupling to gate electrodes even if a system of interest does not have any damping source [27,28]. The capacitive coupling also modifies the charge propagation mode [14,17,26]. Furthermore, crosstalk between high-frequency charge excitation lines and gate electrodes induces unwanted excess charge excitations around the gates.

Here, we present a way to avoid these side effects. The degree of the coupling between a low-dimensional electron system and a gate electrode depends on the high-frequency response of charge carriers in the gate electrode. Since the charging time of a gate electrode is determined by the RC time constant, the coupling can be suppressed by replacing the metal commonly used for a gate electrode with a highly resistive one, for which we choose zinc oxide (ZnO) in this work. The degree of coupling in the GHz range can be evaluated from the velocity of edge magnetoplasmons (EMPs), which are collective charge oscillations propagating in ECs. Since the gate screening of the electric field in EMPs reduces the EMP velocity by more than one order of magnitude [14,17], the velocity is a good measure of the coupling strength. We compare the EMP velocity in three InAs QH systems with different electromagnetic environments—one has a thin ZnO top gate, another has a normal metal (Ti/Au) top gate, and the other does not have a gate. We show that the EMP velocity in the ZnO gate sample is one order of magnitude larger than that in the Ti/Au gate sample and almost the same as that in the ungated sample. This indicates that the gate screening effect is suppressed in the ZnO gate sample. The ZnO gate allows us to measure the velocity of unscreened EMPs for different values of the carrier density $n$. From the EMP velocity as a function of the Landau level filling factor $\nu$ and $n$, we discuss the shape of the edge potential in an InAs quantum well. Our results suggest that using ZnO as gate electrodes is useful for improving the coherence of high-frequency charges in ECs and also for investigating the edge structure.

We used InAs/Al$_{0.7}$Ga$_{0.3}$Sb samples. The two-dimensional electron gas is formed in the 15-nm-wide InAs quantum well with its center located 42.5 nm below the surface. The low-temperature mobility is about $3 \times 10^4$ cm$^2$/Vs at an as-grown electron density of about $5 \times 10^{11}$ cm$^{-2}$. Figure 1(a) schematically shows the structure of a gated sample. The sample edges were defined by mesa etching. After depositing Ti/Au (10/180 nm) for ohmic contacts, the surface was covered with a 20-nm-thick Al$_2$O$_3$ insulating layer by atomic layer deposition (ALD). For the sample with the normal metal gate, the Ti/Au (15/280 nm) top gate was deposited on the Al$_2$O$_3$ layer together with a high-frequency injection gate for EMP excitations. The injection gate has a $10 \times 10$ μm$^2$ overlap with the mesa.. For the sample with the ZnO gate, Al 5%-doped ZnO (20 nm) was formed by ALD [29]. The sheet resistivity of the ZnO film is about $10^5$ Ω/□. After defining the gate area by selectively etching the ZnO layer, a Ti/Au injection gated was formed. For the gated samples, the top gate is spatially separated from the injection gate and the ohmic contacts by 10 μm. We carried out high-frequency transport measurements to obtain the EMP velocity. An EMP pulse is excited by applying a voltage step to the injection gate. It propagates along the ECs and is detected as a time-dependent current on the detector ohmic contact located downstream of the ECs [24]. From the time delay of the current and the path length, the EMP velocity is determined. The path length is 640 μm in the ZnO gate and ungated samples, while it is 300 μm for the Ti/Au gate sample. Magnetic field *B* up to 10 T was used. All measurements were performed at 1.5 K.

Figure 1(c) shows the current on the detector ohmic contact as a function of time and *B* in the ZnO gate sample at the gate bias $V_G = 0$ V. The time origin $t = 0$ is set at the time of the current peak at $B = 0$ T, where charges propagate as two-dimensional plasmons and arrive at the detector without detectable delay [Fig. 1(b)]. The delay generally increases with *B*, and oscillations with minima at integer $\nu = hn/eB$ are superimposed on the trend, where *h* and *e* are Planck's constant and electron charge, respectively. The amplitude and width of the current peak also oscillate with ν. Note that the small bump of the current around $t = 0$ at $B = 10$ T [Fig. 1(b)] is a result of direct crosstalk between the injection and detection high-frequency lines.

In Fig. 2(a), the EMP velocity in the ZnO gate sample is plotted as a function of ν. In the figure, the longitudinal resistance $R_{xx}$ measured by a standard low-frequency lock-in technique is also included. The EMP velocity is on the order of $10^6$ m/s. It oscillates in antiphase with $R_{xx}$, and its peak value increases approximately linearly with ν. Before discussing this behavior, we compare the EMP velocity to that in the ungated sample [Fig. 2(b)] and the Ti/Au gate sample [Fig. 2(c)]. The filling-factor dependence and the value of the velocity in the ungated sample are almost the same as those in the ZnO sample. In the Ti/Au gate sample, on the other hand, the value is more than one order of magnitude smaller than that in the other samples in the whole measured ν range. It is worth noting that the $R_{xx}$ vs ν traces of the three samples are similar, indicating that the samples were not damaged by the fabrication processes.

It is well established that the smaller velocity in a sample with a metal gate is due to the gate screening effect [14,17]. Since the velocity of EMPs is determined by the Coulomb restoring force of displaced charges, the screening of electric field in EMPs slow them down. Conversely, the correspondence of the velocity in the ZnO gate and ungated samples indicates that the screening effect is suppressed in the ZnO gate sample. The difference between Ti/Au and ZnO is their resistivity. We discuss effects of the resistance of a gate on the EMP transport based on a distributed-element circuit model. Within a circuit representation, a chiral EC in a QH state with $R_{xx} = 0$ can be modeled as a channel-ground transmission line with the channel impedance $Z = \sigma_{xy}^{-1}$, where $\sigma_{xy}$ is the Hall conductance [inset of Fig. 3(a)] [30]. The channel capacitance $C_{ch}$ represents the self capacitance arising from the Coulomb interaction in the EC ($C_{ch}$ depends on the edge structure as discussed below) [31]. A resistive gate can be included as a parallel capacitance $C_g$ with a series resistance $R_g$, where $C_g$ represents geometric capacitance between the EC and the gate, and $R_g$ represents the resistance of the gate per unit length along the EC [32]. The position $x$ and time $t$ dependent potential in the transmission line can be expressed by $V(x,t) = \frac{\rho_{ch}}{C_{ch}} = \frac{\rho_g}{C_g} + R_g \frac{\partial \rho_g}{\partial t}$, where $\rho_{ch}$ and $\rho_g$ are the charge on $C_{ch}$ and $C_g$, respectively. Furthermore, continuity equation $\frac{\partial}{\partial t}(\rho_{ch} + \rho_g) + \frac{\partial I}{\partial x} = 0$ holds, where $I(x,t) = \sigma_{xy} V(x,t) \propto \exp[i(\omega t - kx)]$ with $k$ and $\omega$ the wave number and angular frequency, respectively, is the EMP current. Solving these equations leads to

$$\left(C_{ch} + \frac{C_g}{1+i\omega\tau_g}\right)\omega = \sigma_{xy} k, \quad (1)$$

where $\tau_g = R_g C_g$ corresponds to the time constant for charging the gate electrode. In the $\tau_g \to 0$ limit, Eq. (1) gives the group velocity, $v_{sc} = \partial\omega/\partial k = \sigma_{xy}/(C_g + C_{ch})$, which corresponds to the velocity of screened EMPs. In the opposite limit ($\tau_g \to \infty$), the group velocity becomes $v_{unsc} = \sigma_{xy}/C_{ch}$, which corresponds to the velocity of unscreened EMPs. This explains why the EMP velocity in a gated sample changes from screened to unscreened as the resistance of the gate increases.

To see the behavior at intermediate $\tau_g$, we discuss the implication of Eq. (1) in more detail. Eq. (1)

gives two EMP modes:

$$\omega_\pm = \frac{r\omega_{sc}}{2}\left\{1 + \frac{i}{\omega_{sc}\tau_g} \pm \sqrt{\left(1 + \frac{i}{\omega_{sc}\tau_g}\right)^2 - \frac{4i}{r\omega_{sc}\tau_g}}\right\}, \tag{2}$$

where $\omega_{sc} = \sigma_{xy}k/(C_{ch} + C_g)$ and $r = (C_g + C_{ch})/C_{ch}$. The real and imaginary parts of $\omega_\pm/\omega_{sc}$ are plotted as a function of $\omega_{sc}\tau_g$ in Figs. 3(a) and (b). For the plot, we used $r = 20$, which roughly corresponds to the ratio of the EMP velocity at $\nu = 2$ in the ungated and Ti/Au gate samples. The plots indicate that, when $\omega_{sc}\tau_g \gg 1$, the $\omega_+$ mode can propagate ($\text{Re}[\omega_+] \gg \text{Im}[\omega_+]$) with $\text{Re}[\omega_+] \approx r\omega_{sc} = \sigma_{xy}k/C_{ch} \equiv \omega_{unsc}$, while the $\omega_-$ mode is diffusive ($\text{Re}[\omega_-] \ll \text{Im}[\omega_-]$). For $\omega_{sc}\tau_g \ll 1$, on the other hand, only the $\omega_-$ mode can propagate, and $\text{Re}[\omega_-]$ approaches $\omega_{sc}$. Around $\omega_{sc}\tau_g \sim 1$, the imaginary part is comparable to the real part for both $\omega_\pm$ modes. In this regime, EMPs dissipate quickly. Accordingly, as $\omega_{sc}\tau_g$ is increased, the observable EMP mode changes from $\omega_-$ ($\sim \omega_{sc}$) to $\omega_+$ ($\sim \omega_{unsc}$) across the dissipative regime. Measured EMP velocity indicates that the sheet resistivity of the 20-nm-thick ZnO gate, $\sim 10^5$ Ω/□, is high enough to set the system in the $\omega_{sc}\tau_g \gg 1$ regime in the GHz range. On the other hand, the sheet resistivity of the Ti/Au (15/280 nm) gate, $\sim 10^{-2}$ Ω/□, is low enough for $\omega_{sc}\tau_g \ll 1$. It is worth noting that a fine and thinner metal gate, which is often used for mesoscopic systems, has higher resistance. Such a gate could cause non-negligible dissipation of electrons propagating in adjacent ECs.

The ZnO gate allows us to measure the velocity of unscreened EMPs for different values of $n$. Figures 4(a) and (b) show the EMP velocity and $R_{xx}$, respectively, for three values of $n$ as a function of $\nu$. At $n = 5.1 \times 10^{11}$ cm$^{-2}$, the velocity oscillations in antiphase with $R_{xx}$. As $n$ is decreased, the velocity around integer $\nu$ decreases and, at the same time, $R_{xx}$ at integer $\nu$ increases because of lowered mobility.

We discuss the profile of the edge potential in the InAs quantum well from the observed behavior of the EMP velocity. Dispersion of the unscreened EMP contains information on the transverse width of the EMP mode $w$ [33],

$$\omega = \frac{\sigma_{xy} k}{C_{ch}} = \frac{\sigma_{xy} k}{\epsilon}\left(\ln\frac{2}{kw} + c\right), \tag{3}$$

where $\epsilon$ is the effective dielectric constant and $c$ is a constant depending on the potential slope at the sample edge ($c = 1$ for hard-wall edge potential). The reduction of the velocity with increasing $R_{xx}$ suggests that finite conductivity of the bulk 2DES increases $w$. This is expected for a system with hard-wall edge potential [33,34]. In such a system, when $R_{xx} \sim 0$, $w$ is proportional to the ratio of the Coulomb energy to the energy gap between Landau levels. When the bulk conductivity becomes finite, the EMP mode penetrates into the bulk 2DES, leading to the increase in $w$. In a system with soft-wall edge potential like GaAs, on the other hand, $w$ is primarily determined by the potential shape, and the EMP velocity is hardly affected by the bulk conductivity [17,35]. These discussions indicate that the edge potential in InAs is sharp, consistent with the estimation by dc transport measurements [36].

In summary, we showed that the high-frequency coupling between QH ECs and a gate electrode can be suppressed by replacing Ti/Au commonly used for the gate with a highly resistive ZnO thin film. We demonstrated this by showing that the velocity of EMPs in the ZnO gate sample is almost the same as that of unscreened EMPs in the ungated sample. From the measured ν and $n$ dependence of the EMP velocity, we also showed that the edge potential in an InAs quantum well is hard, differing from that in a GaAs quantum well. The application of the ZnO gate is not limited to QH systems. Application of our idea—using highly-resistive gates to suppress side effects of gate electrodes—is not limited to QH systems. It would be useful for reducing the dissipation and investigating unscreened properties of high-frequency charges in a variety of low-dimensional systems including topological materials.

**Acknowledgements**

We thank H. Irie for valuable discussions and H. Murofushi for sample fabrication.

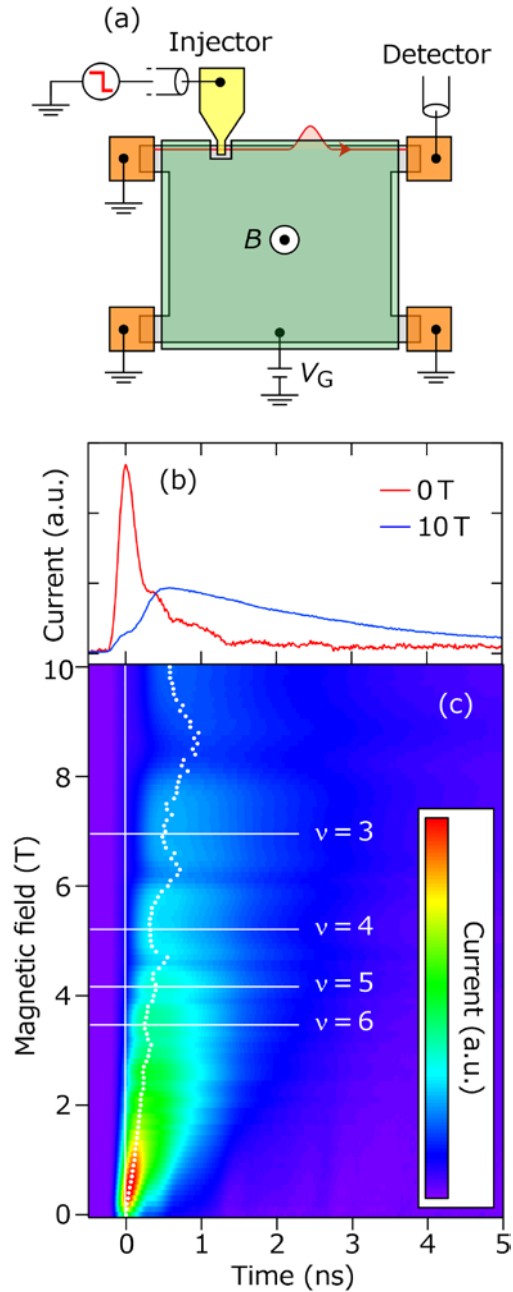

FIG. 1. (color online) (a) Schematic of a sample with a ZnO top gate. Four ohmic contacts (orange) were placed at the corners of the square InAs/Al$_{0.7}$Ga$_{0.3}$Sb mesa. A Ti/Au injection gate (yellow) and a ZnO gate (green) were patterned on an Al$_2$O$_3$ insulating layer. High-frequency lines are connected to the injection gate and the ohmic contact located downstream of chiral ECs. (b) Current in the ZnO gate sample for $V_G = 0$ V at $B = 0$ and 10 T as a function of time. (c) Color-scale plot of the current as a function of time and magnetic field. Dots represents the time for the current peaks. Vertical and horizontal lines represent $t = 0$ and magnetic fields for integer fillings.

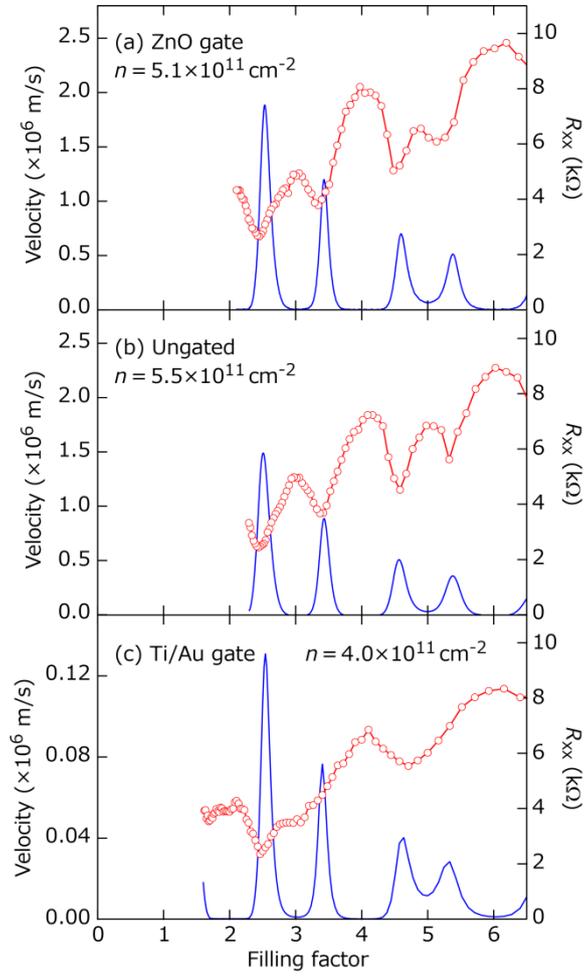

FIG. 2. (color online) EMP velocity (red circles) and $R_{xx}$ (blue line) as a function of the filling factor for the ZnO gate sample (a), the ungated sample (b), and the Ti/Au gate sample (c).

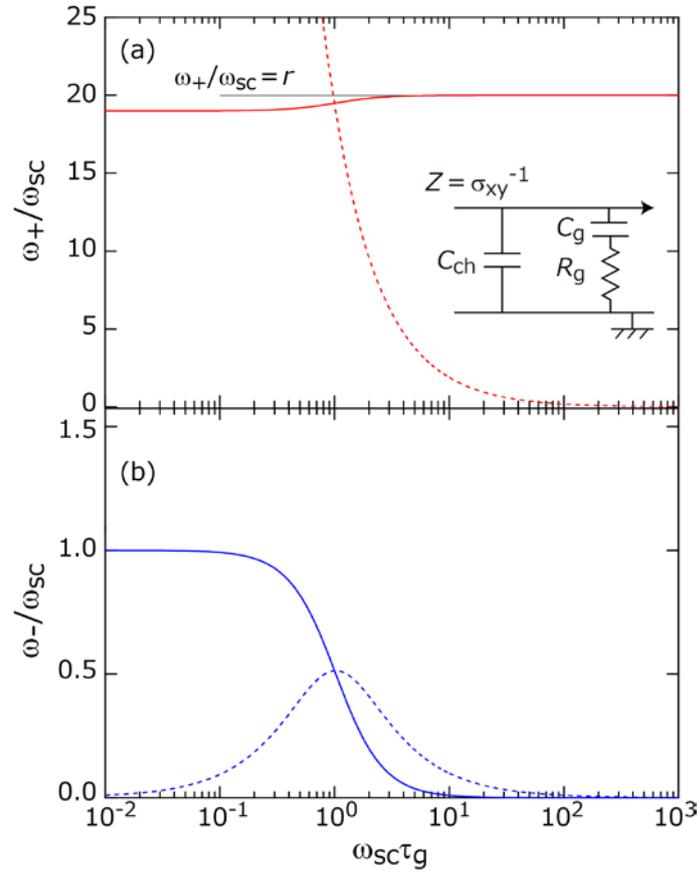

FIG. 3. (color online) (a) and (b) Real part (solid trace) and imaginary part (dotted line) of the $\omega_{\pm}/\omega_{sc}$ as a function of $\omega_{sc}\tau_g$. Inset of (a) represents the circuit model of the EMP transmission line.

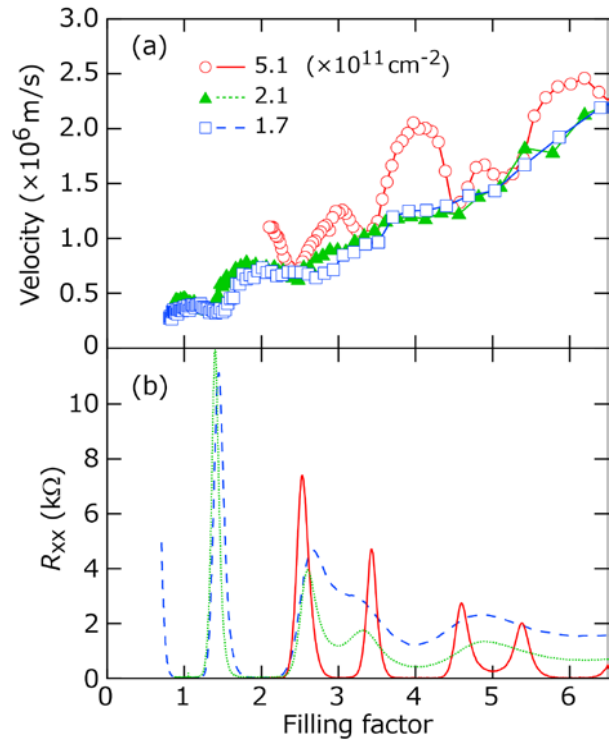

FIG. 4. (color online) EMP velocity (a) and $R_{xx}$ (b) as a function of filling factor for three values of the carrier density controlled by the ZnO gate.